\documentclass{article}

\usepackage{arxiv}

\usepackage[utf8]{inputenc} 
\usepackage[T1]{fontenc}    
\usepackage[hypertexnames=false]{hyperref}      
\usepackage{url}            
\usepackage{booktabs}       
\usepackage{amsfonts}       
\usepackage{nicefrac}       
\usepackage{microtype}      
\usepackage{graphicx}
\usepackage{doi}
\usepackage[sort&compress,square,comma,authoryear]{natbib}
\usepackage{multirow}
\usepackage{tabularx}
\usepackage{xurl}
\usepackage{appendix}
\usepackage{makecell}
\usepackage{placeins}

\graphicspath{ {./images/} }

\title{Preferences and Attitudes towards Debt Collection: \protect\\ A Cross-Generational Investigation}

\author{ {Minou Goetze}\\
	Psychology School, Faculty of Business \& Media\\
	Fresenius University of Applied Sciences\\
	Hamburg, Germany\\
	\texttt{minou.goetze@hs-fresenius.de} \\
	\And
	{Christina Herdt} \\
	PAIR Finance GmbH\\
	Berlin, Germany\\
	\texttt{christina.herdt@pairfinance.com} \\
	\And
	{Ricarda Conrad} \\
	PAIR Finance GmbH\\
	Berlin, Germany\\
	\texttt{ricarda.conrad@pairfinance.com} \\
	\And
	{Stephan Stricker} \\
	PAIR Finance GmbH\\
	Berlin, Germany\\
	\texttt{stephan.stricker@pairfinance.com} \\
}

\date{March 9, 2023}

\renewcommand{\undertitle}{}

\begin{document}
\maketitle

\begin{abstract}
Preliminary research indicated that an increasing number of young adults end up in debt collection. Yet, debt collection agencies (DCAs) are still lacking knowledge on how to approach these consumers. A large-scale mixed-methods survey of consumers in Germany (\textit {N} = 996) was conducted to investigate preference shifts from traditional to digital payment, and communication channels; and attitude shifts towards financial institutions. Our results show that, indeed, younger consumers are more likely to prefer digital payment methods (e.g., Paypal, Apple Pay), while older consumers are more likely to prefer traditional payment methods such as manual transfer. In the case of communication channels, we found that older consumers were more likely to prefer letters than younger consumers. Additional factors that had an influence on payment and communication preferences include gender, income and living in an urban area. Finally, we observed attitude shifts of younger consumers by exhibiting more openness when talking about their debt than older consumers. In summary, our findings show that consumers’ preferences are influenced by individual differences, specifically age, and we discuss how DCAs can leverage these insights to optimize their processes.
\end{abstract}


\section{Introduction}
According to a survey conducted by the Bundesverband Deutscher Inkasso-Unternehmen e.V. (BDIU), an increasing number of young adults end up in debt collection \citep{bundesverband}. On top of that, 38\% of DCAs reported that recovery (i.e., the amount paid back by debtors) was lower among young-age cohorts compared to older-age cohorts. These concerning observations give rise to the question of whether classic debt collection strategies which are efficient among older cohorts are appropriate in addressing younger cohorts. For instance, young consumers are known for their affinity towards digital communication in private settings \citep{forgaysetal14}. Do these findings translate, however, to the debt-collection context? Overall, little is known about the payment and communication preferences of consumers in the debt-collection context as well as their attitudes towards DCAs, despite the important implications that one’s attitudes and preferences may have on the relationship between consumers, creditors and DCAs, as well as on recovery statistics. A plethora of questions arise from these circumstances that DCAs need to challenge in order to gain a better understanding of consumers of various age groups, such as which communication channel do young/old consumers prefer, at which time should the communication happen or which payment methods should be offered. All these factors—and more—ought to be considered if one wants to interact with consumers of different age cohorts in an efficient and consumer-oriented manner.
 
Hence, the main purpose of this study is to gain first insights into age-dependent differences in communication and payment preferences as well as attitudes towards DCAs and being indebted. To do so, we conducted a mixed-methods survey in Germany. By asking for both qualitative judgments on debt collection and quantifiable measures (e.g., past touch points with debt collection and preferences in communication and payment methods), we aim to derive information on how the debt collection process can be optimized for consumers of all ages based on their preferences.

\section{Theoretical Background}
\label{sec:theoretical_background}

\subsection{COM-B Model}

To gain a better understanding of the role age plays in a debt collection context, we used a theoretical framework that was developed to implement behavioral interventions, namely the COM-B system \citep{michieetal11}. According to the COM-B system, a certain behavior is only expressed if a consumer has the necessary capability, opportunity, and motivation to perform such a behavior. In the debt collection context, the desired behavior is a reaction (and eventually payment) by a consumer following a contact point, which was identified as one of the most critical steps in debt collection \citep{ghaffarietal21}. To understand more about possible drivers of reaction behavior, we mapped the three dimensions of the COM-B (Capability, Opportunity, Motivation) onto possible interventions that can be realized by DCAs. Capability refers to an individual’s capacity to engage in the target behavior. In the debt collection context, capability translates to one’s financial ability to pay—an important predictor for the reaction behavior of debtors \citep{ghaffarietal21}. While DCAs cannot influence the financial means of consumers, they can decide what payment methods to offer to consumers and thereby increase their capability to react. Opportunity is defined as all external factors that make the target behavior possible. One way DCAs can influence the opportunity of a consumer to react is by choosing an appropriate communication channel and timing. Motivation is defined as cognitive processes that energize and direct behavior. In the debt collection context, motivation refers to the consumer’s willingness to pay their debt \citep{ghaffarietal21}, which includes factors such as consumer’s attitudes and beliefs about debt. Taking previous applications of the COM-B model into account, behavioral interventions along the three dimensions depend on the age group under investigation \citep{willmottetal21}, \citep{tayloretal16}. Therefore, the present study investigates preferences for payment methods (i.e., capability), preferences for communication (i.e., opportunity), and attitudes towards DCAs (i.e., motivation) by specifically focusing on the role of age. 

\subsection{Payment Methods}

Researchers have extensively investigated the relationship between payment methods, consumer behavior and characteristics. A study by \cite{eschelbachetal22} has shown that despite some clear shifts compared to previous years, such as an increase in the transaction share of the e-payment method Paypal of 2.4\% from 2020 to 2021, German consumers still tend to rely on rather traditional payment methods (i.e., cash, manual transfer) compared to more modern payment instruments (e.g., Apple Pay). More precisely, cash payments were the most frequently used means of payment with a share of 58\%, followed by payments by debit card (23\%), credit card (6\%) and by direct debit/credit transfer (4\%). E-payment methods (e.g., Paypal, Klarna) were found to account for 5\% and mobile payment methods (e.g., Apple Pay) for 2\% of all transactions \citep{eschelbachetal22}. 

As for age-dependent preferences regarding digital payment methods, findings demonstrate that online manual transfer was mostly used by consumers ranging from 16 to 29 years, whereas this was the least preferred payment method for consumers aged between 46 and 60 years, who preferred cash payments the most. Credit transfers, credit cards, cheques and mobile banking were mostly used by consumers between 30 and 45 years of age \citep{camilleriagius21}. Similarly, preliminary findings of another study show that older consumers were less likely to pay via smartphone, smartwatch, or biometric technology \citep{klarna21}. Contrary to older age groups, it was even found that consumers between 18 and 30 years of age were rather skeptical of traditional financial institutions and preferred to conduct their banking and financial business online \citep{mondres19}. These findings are in line with research investigating age-dependent preferences for mobile payments \citep{agardialt22}. In addition, their results revealed that for consumers aged between 43 and 57 years, perceived ease played a critical role in perceived usefulness of mobile payments, while for consumers aged up to 27 years, financial and privacy risks turned out to be insignificant regarding mobile payment. 

However, while past research has provided first evidence on age-dependent differences in payment preferences, little is known about more recent payment methods, such as Paypal and Apple Pay. On these grounds, a systematic investigation on age differences across the most modern online payment methods needs to be undertaken, which is further extended to a debt collection context.

\subsection{Communication}

Past studies have shown that communication of businesses plays a key role in influencing consumer behavior \citep{kusaetal20}, \citep{mihart12}. Over the past decades, digital communication channels such as email and text messaging rose to popularity when improvements in networking and cellphone technology brought digital mediums to a broad consumer base \citep{lipiäinen14}. In recent years, an increasing number of online channels such as live chats and social media, entered the market as new communication tools \citep{edosomwanetal11}, \citep{forrester20}. Previous research has shown that businesses interacting with customers in accordance with their communication preferences achieved better business results \citep{forrester20}. Such findings sparked a novel interest of companies to better understand their consumers’ communication preferences. Consequently, results of a survey revealed that 83\% of global consumers preferred emails when receiving communication from businesses compared to text messaging \citep{twilio19}. Nevertheless, when receiving urgent communications from businesses, text messages were more than twice as popular compared to email communication. The top three key attributes of consumers’ preferred communication channels included convenience (50\%), reliability (45\%), and speed (41\%), which reflects the evolution to a digitally fast-paced world \citep{cmocouncil19}. 

 Regarding age differences, numerous studies show that especially adolescents and young adults identified text messaging as their preferential form of contact when compared with instant communication channels, such as email or talking over the phone (e.g., \cite{faulknerculwin05}, \citep{madellmuncer07}, \citep{pierce09}). In line with that, SMS text messaging (compared to email) showed to be more salient to the recipient (\citep{danaheretal15} and was found to be linked to higher opening and click through rates \citep{muenchbaumel17}.

For adults older than 66 years, however, in-person and written communication instead of technologically mediated communication was preferred \citep{yuanetal16}. In addition to communication channels, few studies investigated consumers preferences concerning timing of communication. Consistent results indicate that emails are more likely to be opened in the morning, rather than later in the evening (e.g., \cite{meyer22}). Regarding online purchases though, order rates were the highest for emails sent out by 4 p.m. \citep{meyer22}. The above-mentioned studies investigated communication preferences of consumers irrespective of the specific industry that is sending outbound communication. Thus, it is unclear how these findings extend to the debt collection industry, particularly with respect to age-related differences.

\subsection{Attitudes}
Recently, a trend was observed on TikTok, where primarily young adults showed off their overdrawn Klarna (a buy-now-pay-later service provider) accounts under the hashtag “\#klarnaschulden”. Given the high media coverage of the trend, it appears attitudes towards indebtedness may be undergoing a shift in Germany, with younger consumers being more open-minded about these topics. Such an attitude shift could have detrimental economic effects though, not only for the consumers and the creditors but the debt collection industry, which needs to adjust to this growing consumer pool \citep{benderbreuer11}. A cross-sectional study among Swedish citizens that assessed participants’ attitudes towards being indebted and their level of uncomfortableness with debt illustrates the importance of one’s attitudes in the context of indebtedness \citep{almenbergetal21}. The authors found that “being uncomfortable with debt is transmitted from parents to children” as well as that “discomfort with debt [...] [is] declining over time” (p. 15). Most importantly, those who feel uncomfortable with debt were found to “have considerably less debt” (p. 12). Being young and generally comfortable with debt could therefore lead to an early onset of accepting one’s indebtedness and, consequently, a long-term career in debt collection could become more likely \citep{benderbreuer11}. Based on interviews with US college students, \cite{zerqueraetal16} developed a continuum of attitudes towards debt, which reflects heterogenous attitudes among young consumers. Their continuum describes three main levels of attitudes: debt-averters were described as those who actively circumvent debt altogether, willing to sacrifice comfort and living conditions. The debt-averters were mainly influenced by observing others' negative experiences in being indebted. In contrast, debt-intermediates were characterized by the belief that some debt was necessary while overall trying to minimize the amount of debt carried out. Their key influences were negative experiences with educational and financial institutions. Lastly, debt-acceptors considered debt to be a normal part of college life and were willing to be indebted to maintain comfort. Their key influences were identified as their needs and preferences. While this continuum provides valuable insights into the different attitudes and mindsets under which young consumers may or may not accumulate debt, the findings cannot be generalized due to the specific demographic and financial context (accumulating debt during college-life) under investigation and the lack of a comparison between older and younger consumers \citep{zerqueraetal16}. 

Despite the observed increase in comfort with debt in younger consumers, negative attitudes towards DCAs seem to prevail. For instance, \cite{jalonentakala18} asked participants aged between 14 and 71 years to describe DCAs in three adjectives. The results of their research indicate a general aversion towards DCAs as the most common responses were of negative sentiment only, namely expensive, greedy, extortionate, uncompromising, threatening, and frightening \citep{jalonentakala18}. 

Overall, we see an interest in attitudes towards being indebted as there are possible implications for one's economic decision making as well as economic effects on the industry. Nevertheless, prior studies often do not account for age-effects, and it is unknown whether the observed effects apply to German consumers as well. Further, it is important to consider not only the consumers’ attitudes towards being indebted but also towards DCAs. As mediators between creditors and consumers, DCAs need to understand and adjust to the consumers to improve the customer-journey and increase recovery.

\section{The Study}

The present study focuses on the three key points. First, we investigated the impact of age on preference for payment methods. Available payment methods differ in terms of traditionality (i.e., how long they have existed in the market) and their level of digitalization (e.g., only available online). Based on preliminary findings \citep{agardialt22}, \citep{camilleriagius21}, we assumed that younger consumers have a stronger preference for less traditional and more digital payment methods than older consumers (H1). Second, we analyzed whether age affects preferences for communication channels. In parallel to payment methods, we assumed that younger consumers prefer to communicate via digital channels rather than non-digital channels (H2). Regarding the timing of communication, we did not have specific hypotheses regarding age differences. Third, we investigated age differences across attitudes towards being in debt and DCAs. Previous literature indicates that consumers hold predominantly negative attitudes towards DCAs \citep{jalonentakala18}. Based on recent observations of the “\#klarnaschulden” trend though, we challenge this notion and hypothesize that younger consumers will have a more positive attitude towards DCAs and being in debt than older consumers (H3).

\section{Methodology}

\subsection{Subject pool}

The convenience sample was sampled through different channels to obtain a sufficient amount of data. Participants were recruited via the Applied University Fresenius Hamburg and the website Clickworker. In total 1000 participants took part in the survey. We only included participants aged between 18 and 60 years. Participants aged below 18 or above 60 years were excluded from any analyses (\textit{n} = 4), resulting in a final sample of 996 participants (\textit{M} = 37.2, \textit{SD} = 10.6, 40\% female). No prior experience with debt collection was required to partake in the study. Participation was incentivized by giving away two 50 Euro Amazon vouchers among the participants.

\subsection{Materials and Design}

To investigate our research questions, we set up an online questionnaire, using a mixed-methods design as the survey consisted of quantitative and qualitative components. The questionnaire covered three main topics: 1) payment preferences, 2) communication preferences, and 3) attitudes toward debt collection. Each topic contained multiple questions. Regarding payment and communication preferences, participants were asked to rank proposed payment methods and communication channels from most preferred to least preferred according to their preference. The options to be ranked included most frequently used payment methods and communication channels in e-commerce in Germany. For payment methods, this included manual transfer, manual transfer via Klarna, credit card, Barzahlen\footnote{Barzahlen (translates to “cash payment”) is a payment method in Germany that offers online consumers to pay their invoice in cash at selected retail-partners.}, Paypal, and Apple Pay. For communication channels this included letters, emails, calls, SMS and Whatsapp. Attitudes towards DCAs were mainly gathered qualitatively, though one question was examined quantitatively by selecting a single-choice answer. Three open questions recorded attitudes by asking participants to provide three adjectives in response to each question. Additional questions included trust in email communication and ratings of the content of payment reminder messages. Further, data including age, gender, place of residence (indicated by the postal code), monthly income and rent per square meter were collected.

\subsubsection{Procedure}

Data collection took place between April and September 2022 and was conducted concurrently with recruitment. The questionnaire was administered through the online survey platform Unipark. Before the questionnaire could be accessed, online informed consent had to be provided by clicking a required checkbox. Thereupon, the participant was informed about the estimated completion time of 15 minutes. The participant was advised to fill out the questionnaire in one session since the survey could not be continued if quit throughout the session or taken more than once.
First, participants were informed about the aim of the study and asked to read a general introduction to the purpose of a DCA. Second, questions about participants’ experiences with debt collection were posed. If applicable, reasons for ending up in a debt collection procedure had to be stated. Next, a total of 19 content questions were posed, followed by a basic set of socio-demographic questions (see Appendix for full questionnaire). The three main topics (payment preferences, communication preferences and attitudes towards debt collection) were covered sequentially.

\subsection{Data Analysis}
\subsubsection{Quantitative}

To find out whether age differences have an impact on payment method and communication preferences, we used an ordered logistic regression predicting ranked preferences by age. In order to test the stability of the result, we controlled for participant gender, monthly income, and residence. Urban residence indicates if participants live in a rural or urban area, depending on population density; the variable was created based on zip codes and ranged from 1 being the least urban to 3 being the most urban. For any regression analysis, we excluded observations where the reported monthly income is above the 99th percentile (above 7800 Euros per month), assuming that participants reported their yearly income (\textit{n} = 36). This exclusion procedure left us with a final sample of 872 participants for the regression analysis (\textit{M} income = 2055, \textit{SD} = 1222). For a simplified visual display of age effects, we distinguished between three age groups, which were created based on generations commonly referred to as Generation X (46-60 years, \textit{n} = 220), Y (31-45 years, \textit{n} = 485) and Z (18-30 years, \textit{n} = 220). Note that our design was not experimental and consequently we do not draw any causal inferences but instead report correlational evidence.

\subsubsection{Qualitative}
Regarding the qualitative analysis of the adjectives representing attitudes towards DCAs, we established two exclusion criteria for the responses. First, responses longer than one word (full phrases, sentences). Second, one-word responses of other word classes (mainly nouns and verbs), unless the noun/verb was a close match to its adjective derivative and could be transformed into an adjective. Excluding the responses according to these criteria and any missing values, a total of 7280 responses were classified into negative, positive, and neutral sentiment by the Bert model \citep{guhretal20}. While the model was able to classify each token, a manual review by a native speaker was performed to ensure correct classification. In a next step, an ordinal regression was performed. Here, we used participants’ age (continuous variable), gender and prior experience with DCAs (0 = none, 1 = at least once) to predict attitudes towards DCAs measured by the sentiment of the adjectives (negative, neutral, positive). We excluded participants from the regression analysis if they indicated "I do not know" with regards to their experience with DCAs (\textit{n} = 11). Analogous to the quantitative analysis, we distinguish between three age groups commonly referred to as Generation X (46–60 years), Y (31–45 years) and Z (18–30 years) in the descriptive analysis. 

\section{Results}

\subsection{Payment Preferences}

Overall, our results indicate that Paypal was the most and Apple Pay the least preferred payment method. Specifically, Paypal was rated most frequently as the preferred payment method (36.55\%), while other payment methods such as manual transfer (26.61\%), manual transfer via Klarna (5.52\%), credit card (9.14\%), or Apple Pay (2.31\%) were rated less frequently as the most preferred option.\footnote{Payment method “Barzahlen” (19.88\%) was excluded from all analyses, since it was most likely misunderstood by participants as referring to standard cash payments.} 

\begin{figure}[ht]
	\centering
	\caption{Differences in preferences for payment methods across age groups. Payment methods are ranked on a scale ranging from 0 (least preferred) to 6 (most preferred).}
	\includegraphics[width=16cm, height=12cm]{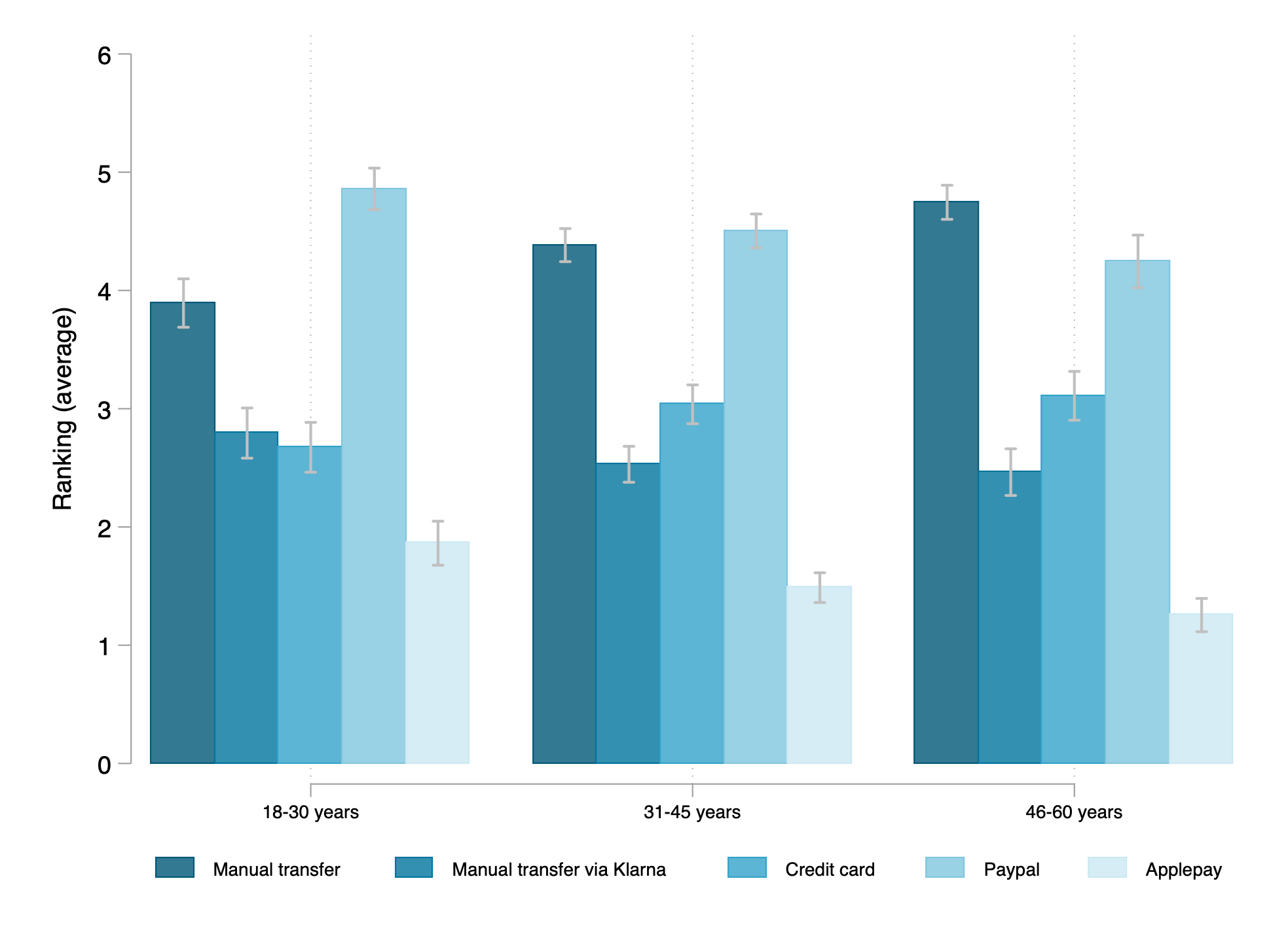}
	\label{fig:payment_preferences}
\end{figure}

When analyzing differences across age groups in terms of preference for payment method, the results support our H1 and show that there are substantial age-dependent differences (see Figure~\ref{fig:payment_preferences}). On average, Paypal was the most preferred payment method by participants aged between 18-30 years (\textit{M} = 4.86) and 31-45 years (\textit{M} = 4.5), while manual transfer was the most preferred by those aged between 45-60 years (\textit{M} = 4.75). Specifically, we found that older participants were significantly more likely to prefer manual transfer $(OR = 1.03, z = 5.47, p < .001)$ than younger participants. On the other hand, younger participants significantly preferred using Paypal $(OR = 0.97, z = -5.21, p < .001)$ and Apple Pay $(OR = 0.97, z = -4.70, p < .001)$. In addition, we found effects of gender, income, and living in an urban area on payment method preference. In particular, women were more likely to prefer manual transfer via Klarna than men $(OR = 1.78, z = 4.42, p < .001)$ and men were more likely to prefer credit card payments $(OR = 0.61, z = -3.87, p < .001)$. Regarding income, a higher income was associated with a higher preference for credit cards $(OR = 1.00, z = 2.51, p = .012)$, Paypal $(OR = 1.00, z = 2.39, p =.017)$ and Apple Pay $(OR = 1.00, z = 2.81, p = .005)$. Living in an urban area was a significant predictor for preferring credit cards $(OR = 1.25, z = 2.59, p = .009)$ and Apple Pay $(OR = 1.23, z = 2.71, p = .007)$. For an explorative analysis of geographical influences on payment preferences, see Appendix (Figure~\ref{fig:geography_payments}).

\subsection{Communication Preferences}

Overall, regarding communication channel preferences our results indicate that letters were the most and phone calls the least preferred channels. Specifically, letters were rated most frequently as the preferred channel (49.10\%), closely followed by emails (43.88\%), while other channels such as Whatsapp (4.22\%), SMS (1.71\%) or phone calls (1.10\%) were rated less frequently as the most preferred option.

\begin{figure}[ht]
	\centering
	\caption{Differences in preferences for communication channels across age groups. Communication channels are ranked on a scale ranging from 0 (least preferred) to 5 (most preferred).}
	\includegraphics[width=16cm, height=12cm]{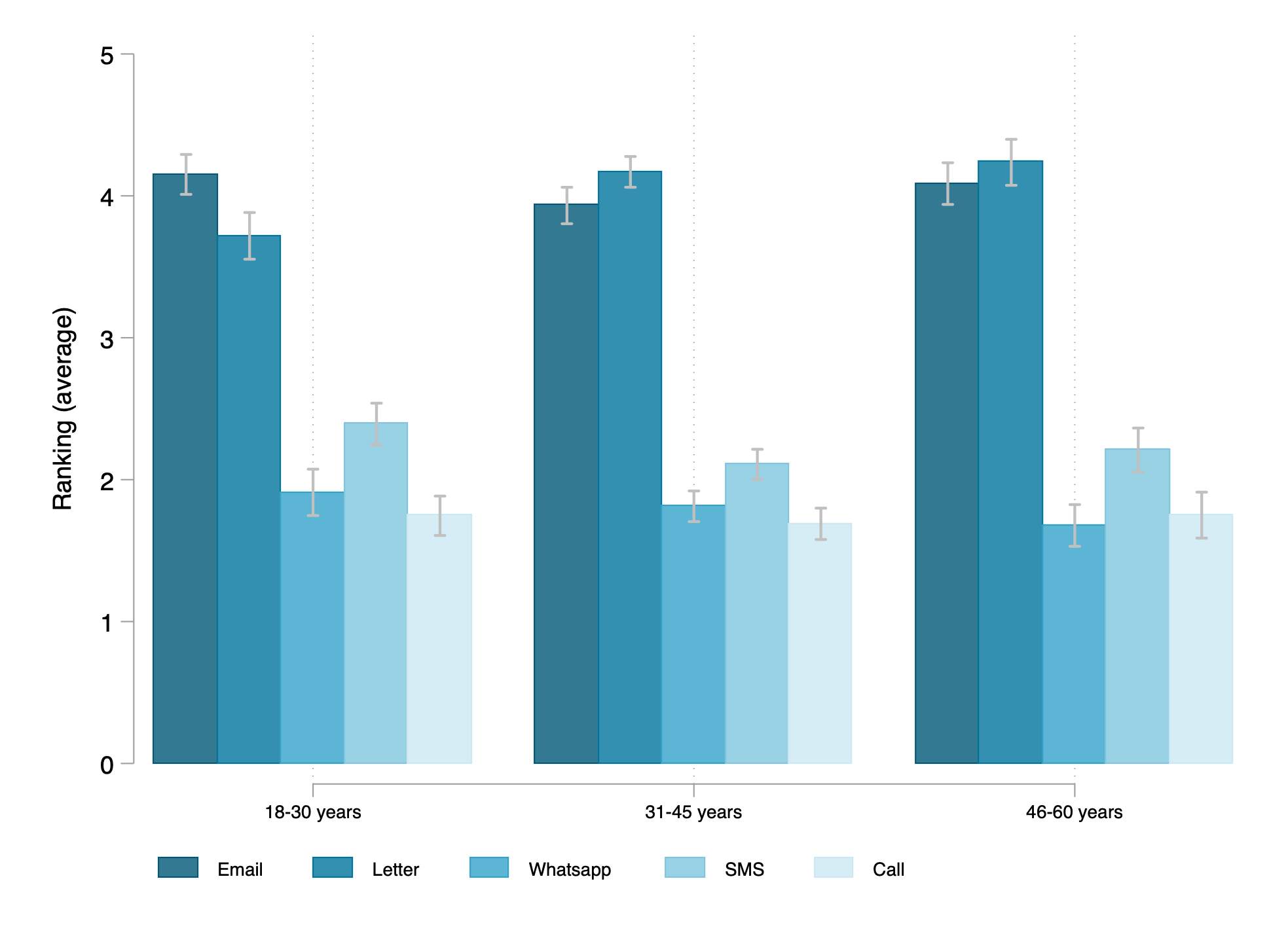}
	\label{fig:channel_preferences}
\end{figure}

When analyzing differences across age groups in terms of preference for communication channels, the results show only significant differences for letters (see Figure~\ref{fig:channel_preferences}). On average, emails were the most preferred channel by participants aged between 18-30 years (\textit{M} = 4.15), while letters were the most preferred by those aged between 31-45 years (\textit{M} = 4.17) and 45-60 years (\textit{M} = 4.24). In line with H2, older participants were significantly more likely to prefer letters than younger participants $(OR = 1.03, z = 5.45, p < .001)$. In addition, we found effects of income and urban residency on channel preference. In particular, a higher income was associated with a preference for Whatsapp $(OR = 1.00, z = 3.32, p = .001)$. Participants living in an urban area were more likely to prefer SMS as a communication channel than participants living in a more rural area $(OR = 1.32, z = 3.14, p = .002)$. There were no significant differences regarding communication channel preferences depending on gender. For an explorative analysis of geographical influences on communication channel preferences, see Appendix (Figure~\ref{fig:geography_channels}).

In an explorative analysis, we investigated time preferences in more detail. Overall, regarding preferences at which time of day participants would like to receive communication from a DCA, our results indicate that noon was the most and evening the least preferred time of day for communication across all age groups. Specifically, noon was rated most frequently as the preferred time (31.12\%), closely followed by morning (30.02\%), while afternoon (24.50\%) and evening (14.36\%) were rated less frequently as the most preferred option. 

\begin{figure}[ht]
	\centering
	\caption{Differences in preferences for time of day across age groups. Communication times are ranked on a scale ranging from 0 (least preferred) to 4 (most preferred).}
	\includegraphics[width=16cm, height=12cm]{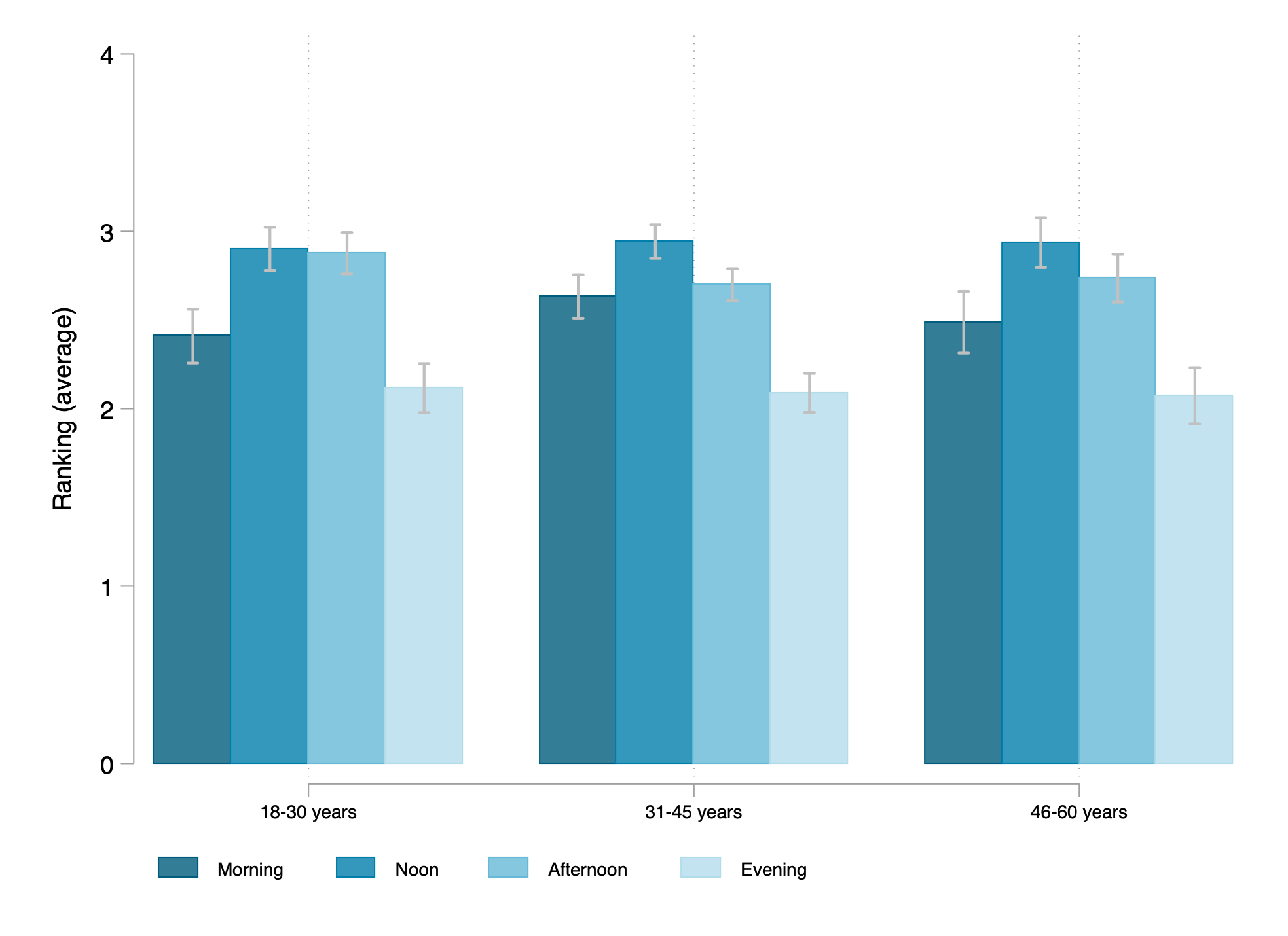}
	\label{fig:time_preferences}
\end{figure}

When analyzing differences in time preferences across age groups, the results indicate that only marginal age-dependent differences can be observed (see Figure~\ref{fig:time_preferences}). Specifically, younger participants were significantly more likely to prefer communication in the afternoon than older participants $(OR = 0.99, z = -2.07, p = .038)$. In addition, we found effects of gender, income, and urban residency on time preference. In particular, males had a stronger preference for receiving communication in the afternoon than females $(OR = 0.71, z = -2.44, p = .015)$. Regarding income, higher income was associated with a preference for communication in the morning $(OR = 1.00, z = 2.72, p = .006)$. Participants living in more rural areas were more likely to prefer communication in the evening $(OR = 0.79, z = -2.46, p = .014)$. For an explorative analysis of geographical influences on communication time preferences, see Appendix (Figure~\ref{fig:geography_timing}).

\subsection{Attitudes}
Overall, we observed differences between participants’ openness to talk about being contacted by a DCA depending on their age. When asked whether they would tell a friend about being contacted by a DCA, 43\% of participants aged between 18–30 years opted for the option “Yes, but only with my close friends”, and only 24\% opted for “No, I would keep it to myself”. Participants aged between 31–45 years however were split between the two responses, with 35\% indicating they would tell their close friends but 34\% indicating they would keep it to themselves. Similar distributions were observed for participants aged 46–60 of whom 38\% indicated they would tell their close friends and 36\% indicated they would keep it to themselves. Particularly, older participants were more likely to indicate that they would keep being contacted by a DCA to themselves $(OR = 1.01, z = 2.13, p = .033)$. In contrast, no effects of gender were observed $(OR = 1.26, z = 1.81, p = .071)$.
 
Regarding participants’ general opinion towards DCAs, most responses expressed a negative sentiment (67.6\%), the most common adjectives being illegitimate, unpleasant, and expensive. Only about one in four adjectives expressed a positive sentiment (27\%), among which necessary, helpful, and useful were most common. Neutral sentiments were the rarest (5.3\%), the most common adjectives being resolute, fast, and assertive. For a more detailed overview of adjectives associated with DCAs, see Figure~\ref{fig:attitudes_1} (for the original German adjectives and their frequencies, see Appendix Table~\ref{tab:adjectives_1}).

\begin{figure}[ht]
	\centering
	\caption{Attitudes towards DCAs expressed via adjectives. Larger font sizes indicate more frequent occurrences (minimum frequency displayed is 15). Adjectives with a positive sentiment are displayed in orange, negative sentiments are displayed in blue and neutral ones are displayed in pink. The figure includes 55.2\% of the responses rated for sentiment.}
	\includegraphics[width=10cm, height=9cm]{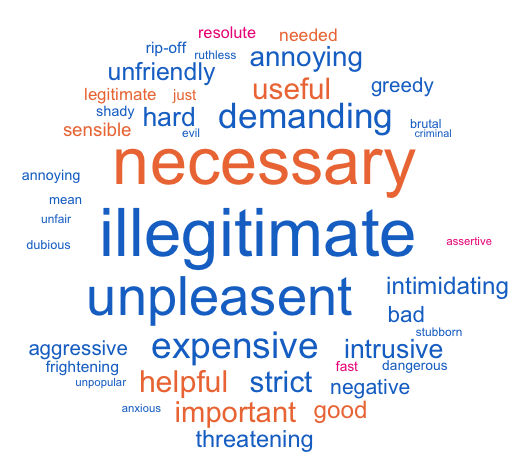}
	\label{fig:attitudes_1}
\end{figure}

\FloatBarrier
Concerning age-effects, the results show that participants aged between 18–30 years were less likely to associate DCAs with a negative sentiment (65.8\%) than participants aged between 31–45 years (68\%) and 46–60 years (70.2\%). In contrast to H3, regression results indicate that age is not a significant predictor for attitudes towards DCAs $(OR = 0.99, z = -1.42, p = .15)$. 
Investigating gender effects regarding participants’ attitudes towards DCAs, we observed substantial differences. The probability to associate DCAs with a positive sentiment was higher for women (31\%) than for men (23\%). Conversely, the probability to associate DCAs with a negative sentiment was lower for women (62\%) than for men (71\%). When analyzing gender differences in attitudes towards DCAs in more detail, regression results indicate significant differences depending on the participants’ gender. Particularly, women were more likely to use adjectives that expressed a positive sentiment to describe DCAs $(OR = 1.45, z = 4.49, p < .001)$.

Concerning prior experience with DCAs, most participants indicated to have never been contacted by a DCA (71\%). 18.1\% of participants had been contacted once and 10.9\% were contacted more than once. Of those who had been contacted once, 11.8\% reported a positive experience while 50.6\% percent reported a negative experience. Similarly, of those who had been contacted more than once, 9.3\% indicated their experience was positive while 60.7\% reported a negative experience. In an ordinal regression analysis, prior experience with DCAs was found to have a significant impact on the adjectives used to describe DCAs. Participants who had been in contact with a DCA at least once were more likely to provide an adjective with negative sentiment compared to participants who had never been in contact with a DCA $(OR = 0.42, z = -8.30, p < .001)$.\footnote{Age and prior experience were not correlated $(r = 0.017, p = 0.34)$}

When participants were asked to describe the characteristics of people who get contacted by DCAs, most adjectives expressed negative sentiments (90\%), the most common responses being poor, unreliable, and unorganized. In contrast, neutral and positive sentiments were less frequently expressed (3.9\% and 6.1\% respectively). For a more detailed overview of adjectives associated with people who get contacted by DCAs, see Figure~\ref{fig:attitudes_2} (for the original German adjectives and their frequencies, see Appendix Table~\ref{tab:adjectives_2}). Prior experience with DCAs was associated with providing more positive adjectives for individuals who were contacted by DCAs  $(OR = 1.67, z = 3.58, p < .001)$. The sentiment of adjectives was not influenced by participants’ gender $(OR = 0.86, z = -1.00, p = 0.31)$ or their age $(OR = 0.98, z = -1.59, p = 0.11)$.

\begin{figure}[ht]
	\centering
	\caption{Attitudes towards people contacted by Debt Collection Agencies expressed via adjectives. Larger font sizes indicate more frequent occurrences (minimum frequency displayed is 15). Adjectives with a positive sentiment are displayed in orange, negative sentiments are displayed in blue and neutral ones are displayed in pink. The figure includes 53.8\% of the responses rated for sentiment.}
	\includegraphics[width=10cm, height=9cm]{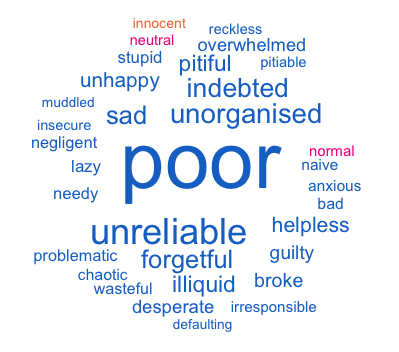}
	\label{fig:attitudes_2}
\end{figure}

\FloatBarrier
\section{Discussion}

The present study provides an extensive overview of age-dependent payment and communication preferences in a debt collection setting and attitudes towards debt collection. Firstly, we demonstrated age differences in terms of preferences for payment methods. As expected, our results revealed a stronger preference for digital payment methods (e.g., Paypal, Klarna, and Apple Pay) among younger consumers and a stronger preference for traditional payment methods (e.g., manual transfer and credit cards) among older consumers. In a similar vein, our findings have shown that younger consumers preferred to receive communications via digital channels (e.g., email, Whatsapp, SMS) over non-digital channels (e.g., letter, call), while older consumers preferred to receive communications via analog communication channels (letters). Thirdly, we have shown that age-linked differences are also reflected in attitudes towards DCAs. In line with our hypothesis, our findings demonstrate that younger consumers are generally more open to talk about indebtedness than older consumers. However, the perception of DCAs and indebted consumers is still negative among all age cohorts.

Previous research supports the notion of age differences in digital affinity. For example, younger citizens were more likely to use online sources to read the news and politically educate themselves than older citizens \citep{boulianneshehata22}, \citep{tanejaetal18}. In line with our findings on communication channels, past studies found that younger individuals use mobile phones more frequently to communicate with friends compared to older people \citep{forgaysetal14}. A higher affinity for innovative payment technologies may be the reasoning for younger consumers to prefer online payment instruments, as this cohort of consumers grew up with the internet and evolving innovative payment methods \citep{kolnhoferetal17}. Similarly, a qualitative study investigating communication preferences of older adults (older than 66 years) via semi-structured interviews revealed stronger preferences for in-person and written communication rather than technologically mediated communication \citep{yuanetal16}. Our results build on these findings and show that older consumers not only prefer analog communication methods in private contexts but also in business contexts while younger consumers seem to prefer digital communication methods in both contexts. By offering digital payment methods and selecting digital communication channels for younger consumers, DCAs can implement behavioral interventions based on the COM-B model and enhance consumers’ capability and opportunity to react.

One possible explanation for cross-generational differences in payment and communication preferences are differences in openness, one of the so-called Big Five personality traits. Past research has found that levels of openness to new experiences (i.e., new payment methods or communication channels) decrease across the lifespan of consumers \citep{mccraeetal99}, \citep{donnellanlucas08}. This decrease in openness could be a potential driver of preferences for more traditional payment methods and communication channels in older consumers and, vice versa, for more digital alternatives in younger consumers. Another explanation for age-dependent differences in terms of digital communication channels can be offered by the link between cognitive load and social media. The attention span of an average reader was found to have decreased from twelve to eight seconds since the rise of new online mediums \citep{microsoft15}, such as social media. A key reason for the narrowing of collective global attention is the overabundance of information as consumers are nowadays balancing many information streams throughout their day. Hence, communications via SMS or Whatsapp messages are likely to be less of a burden due to their short and direct format, at least for consumers—presumably young consumers—that are regularly affected by the daily overload of online input \citep{lorenzetal19}. Regarding time preferences in communication, the results of our study revealed a general preference to be contacted by DCAs around noon and afternoon. Even though findings on age and circadian rhythms \citep{evansetal21} suggest that differences in sleeping patterns could be reflected in time preferences for communication, we did not find any differences regarding time preferences across age groups. One possible explanation for similar time preferences across generations could be the specific business context here. Being contacted by companies (particularly DCAs) outside of regular business hours might be considered inappropriate by consumers and shift time preferences, irrespective of age, to noon and afternoon. 

Concerning attitudes towards DCAs and indebtedness, we have found that the previously observed negative stigma \cite{jalonentakala18} is still prevalent across all age groups. It should be noted that most participants in our sample had never been in contact with DCAs, i.e., most of the attitudes were not based on first-hand experience. Nevertheless, consumers with prior experience with DCAs exhibited more negative attitudes towards DCAs, which holds important implications for the debt collection industry. The dichotomy between the two most common responses “illegitimate” and “necessary” is especially interesting. While many people can rationalize and see the need for DCAs to exist, a great number of people continue to question the legitimacy of the industry. Such attitudes are likely to impact both parties: the motivation to react/pay in debtors and the DCAs, which need to pay attention to communicate transparently and consumer-friendly to justify their mandate to the debtor. When taking age into account, a more nuanced picture emerges. While age was not a significant predictor for attitudes towards DCAs, we still observed a significantly higher openness to talk about being contacted by DCAs in younger participants.

A multitude of factors could explain the observed findings. For instance, an increasing number of industries put emphasis on the customer-journey and consumer-oriented communication—debt-collection being no exception. In the past, the debt-collection process in Germany was less regulated, to the detriment of the consumers. Between 2010 and 2022 multiple laws have been enacted and revised to ensure transparency in the debt collection process, and to cap the fees DCAs are allowed to charge (e.g., \cite{inkasso20}, \cite{recht21}). Besides these legal advancements, some DCAs have started to work with insights from the fields of behavioral science to further optimize the debt-collection experience for consumers \citep{ghaffarietal21}, although this approach is not the industry standard. The predominantly negative attitudes towards DCAs, especially among participants with prior touch-points with DCAs, suggest that there seems to be further room for improvement in the processes of debt collection.

Concerning the greater openness to talk about debt in younger consumers, the process of digitalization and the stark increase in buy-now-pay-later payment services are important factors that should be considered. With the rise of e-commerce and easy access to buy-now-pay-later payment services, young, digitally affine consumers \citep{boulianneshehata22}, \citep{tanejaetal18}, \citep{forgaysetal14} with on-average lower income \citep{bundesamt23} have more opportunities than ever to accumulate debt. While the usage of a buy-now-pay-later service offers financial flexibility, it could have negative consequences for, especially, young consumers who struggle with economical budget management \citep{bundesamt19}. As postponing payments is becoming more common, it is not surprising that consumers’ openness to talk about indebtedness is shifting. While postponing a payment is not equal to defaulting and being indebted, it is plausible to assume that acceptance and openness to talk about debt are affected by the growing commonality of being able to purchase items without the necessary means to pay for them immediately.

In the present study, we elicited preferences of consumers in a hypothetical debt-collection scenario. While this approach allows us to infer information from both debtors and non-debtors, it is questionable whether one's preferences can be translated into reaction behavior in a real debt-collection setting. For instance, young consumers may prefer to be contacted via email, but would they react to an email in the same manner as to a letter? Particularly for DCAs, it would be important to understand how preferences not only translate into reaction behavior, but also affect repayment probabilities. Such insight would be crucial to measure the actual financial benefits of acting in line with consumers’ preferences. With the current design, we cannot draw any causal conclusions, but encourage future research to investigate this topic. Ideally, this would be investigated systematically in a field experiment to understand possible causal links. Similarly, it would be interesting to investigate the link between attitudes towards DCAs and actual repayment probabilities in a field experiment. Additionally, it should be noted here that the current study was conducted with German consumers only. It can be assumed that consumers’ payment and communication preferences vary to some degree across cultures. Thus, one should consider cultural differences before generalizing the present results to other consumer groups.

\section{Conclusion}

The critically observed increase in young debtors poses many threats and questions. Not only the consumers themselves are at risk, as an early onset of indebtedness may lead to long-term financial problems and a higher risk of indebtedness throughout life, but the lower recovery rates among young consumers may negatively affect the economy. As mediators between consumers and creditors, it is important for DCAs to understand how to successfully communicate with this consumer pool. By identifying the consumer’s capability, opportunity, and motivation (as defined by the COM-B model), DCAs can develop strategies to help debtors pay off their debt. The present study is among the first to explore age-dependent preferences in payment methods and communication channels as well as attitudes in the context of debt-collection. Via a mixed-methods survey, we elicited valuable insights into consumers’ preferences of more digital payment methods (e.g., Apple Pay, Paypal) and communication channels, showing that younger consumers prefer digital communication and payment methods in the context of debt-collection. These insights can help financial institutions (i.e., DCAs) in adjusting their communication channels and payment methods and tailoring their processes based on consumer age. Further, based on the finding that prior experience with a DCA increases the probability of having a negative attitude towards DCAs, we encourage DCAs to review their processes and work with insights from the field of behavioral science to improve the customer experience in debt collection, and thereby changing the negative public perception of this industry. Our findings on attitudes show that the perception of DCAs is still predominantly negative. To facilitate efficient communication and improve image concerns, it is in the interest of financial institutions to approach consumers in a way that is tailored to their individual preferences.

\newpage

\bibliographystyle{plainnat}


\begin{thebibliography}{10}
 
    \bibitem[Agárdi \& Alt(2022)]{agardialt22}
    Agárdi, I., \& Alt, M. A. (2022). Do digital natives use mobile payment differently than digital immigrants? A comparative study between generation X and Z. {\em T Electronic Commerce Research}, 1-28. \url{https://doi.org/10.1007/s10660-022-09537-9}
    
    \bibitem[Almenberg et al.(2021)]{almenbergetal21}
    Almenberg, J., Lusardi, A., Säve-Söderbergh, J., \& Vestman, R. (2021). Attitudes Toward Debt and Debt Behavior. {\em The Scandinavian Journal of Economics, 123}(3), 780-809. \url{https://doi.org/10.1111/sjoe.12419}

    \bibitem[Bender \& Breuer(2011)]{benderbreuer11}
    Bender, N., \& Breuer, K. (2011). {\em Junge Menschen und frühe Schulden – finanzielle Handlungskompetenz im Fokus wirtschaftspädagogischer Forschung.} In Exzellenzcluster „Gesellschaftliche Abhängigkeiten und soziale Netzwerke“ (Ed.) Krisen und Schulden – Historische Analysen und gegenwärtige Herausforderungen. Wiesbaden: VS Verlag, 45–62.
    
    \bibitem[Boulianne \& Shehata(2022)]{boulianneshehata22}
    Boulianne, S., \& Shehata, A. (2022). Age differences in online news consumption and online political expression in the United States, United Kingdom, and France. {\em The international journal of press/politics, 27}(3), 763-783. \url{https://doi.org/10.1177/19401612211060271}
 
     \bibitem[Bundesverband Deutscher Inkasso-Unternehmen e.V., 2017]{bundesverband}
     Bundesverband Deutscher Inkasso-Unternehmen e.V. (2017, December 6). {\em Inkasso-Umfrage: Überschuldung von Verbrauchern setzt Gläubigern zu} [Press release]. \url{https://www.verbaende.com/news/pressemitteilung/inkasso-umfrage-ueberschuldung-von-verbrauchern-setzt-glaeubigern-zu-jeder-zehnte-kann-rechnungen-nicht-puenktlich-bezahlen-119338/}
     
    \bibitem[Camilleri \& Agius(2021)]{camilleriagius21}
    Camilleri, S. J., \& Agius, C. (2021). Choosing Between Innovative and Traditional Payment Systems: An Empirical Analysis of European Trends. {\em Journal of Innovation Management, 9}(4), 29-57. 
    
    \bibitem[CMO Council(2019)]{cmocouncil19}
    CMO Council (2019). {\em Critical channels of choice.} Pitney Bowes. \url{https://www.pitneybowes.com/content/dam/pitneybowes/us/en/campaign-pages/cmo-council-report/critical-channels-pb-report-full-data-final-w-commentary.pdf}

    \bibitem[Danaher et al.(2008)]{danaheretal15}
    Danaher, B. G., Brendryen, H., Seeley, J. R., Tyler, M. S., \& Woolley, T. (2015). From black box to toolbox: Outlining device functionality, engagement activities, and the pervasive information architecture of mHealth interventions. {\em Internet interventions, 2}(1), 91-101. \url{https://doi.org/10.1016/j.invent.2015.01.002}
    
    \bibitem[Donnellan \& Lucas(2008)]{donnellanlucas08}
    Donnellan, M. B., \& Lucas, R. E. (2008). Age differences in the Big Five across the life span: evidence from two national samples. {\em Psychology and aging, 23}(3), 558-566. \url{https://doi.org/10.1037/a0012897}

    \bibitem[Edosomwan et al.(2011)]{edosomwanetal11}
    Edosomwan, S., Prakasan, S. K., Kouame, D., Watson, J., \& Seymour, T. (2011). The history of social media and its impact on business. {\em Journal of Applied Management and entrepreneurship, 16}(3), 79-91.
 
    \bibitem[Eschelbach et al.(2022)]{eschelbachetal22}
    Eschelbach, M., Lorek, K., Novotny, J., Pietrowiak, A., \& Seiler, V. (2022). Payment behaviour in Germany in 2021. {\em Deutsche Bundesbank}, 1-58.
     
    \bibitem[Evans et al.(2021)]{evansetal21}
    Evans, M. A., Buysse, D. J., Marsland, A. L., Wright, A. G., Foust, J., Carroll, L. W., ... \& Hall, M. H. (2021). Meta-analysis of age and actigraphy-assessed sleep characteristics across the lifespan. {\em Sleep, 44}(9), 1-19. \url{https://doi.org/10.1093/sleep/zsab088}

    \bibitem[Faulkner \& Culwin(2005)]{faulknerculwin05}
    Faulkner, X., \& Culwin, F. (2005). When fingers do the talking: a study of text messaging. {\em Interacting with computers, 17}(2), 167-185. \url{https://doi.org/10.1016/j.intcom.2004.11.002}

    \bibitem[Forgays et al.(2014)]{forgaysetal14}
    Forgays, D. K., Hyman, I., \& Schreiber, J. (2014). Texting everywhere for everything: Gender and age differences in cell phone etiquette and use. {\em Computers in Human Behavior, 31,} 314-321. \url{https://doi.org/10.1016/j.chb.2013.10.053}

    \bibitem[Forrester Consulting(2020)]{forrester20}
    Forrester Consulting (2020, December).  {\em What Businesses Need To Know About Communicating With Consumers}. Google. \url{https://developers.google.com/business-communications/business-messages/files/google-what-businesses-need-to-know-about-communicating-with-consumers.pdf}

    \bibitem[Gesetz zur Verbesserung des Verbraucherschutzes im Inkassorecht(2020)]{inkasso20}
    Gesetz zur Verbesserung des Verbraucherschutzes im Inkassorecht und zur Änderung weiterer Vorschriften (2020). \url{https://dip.bundestag.de/vorgang/gesetz-zur-verbesserung-des-verbraucherschutzes-im-inkassorecht-und-zur-\%C3\%A4nderung/261560}
    
    \bibitem[Ghaffari et al.(2021)]{ghaffarietal21}
    Ghaffari, M., Kaniewicz, M., \& Stricker, S. (2021). Personalized Communication Strategies: Towards a New Debtor Typology Framework. {\em Psychology and Behavioral Sciences, 10}(6), 256-268.

    \bibitem[Guhr et al.(2020)]{guhretal20}
    Guhr, O., Schumann, A. K., Bahrmann, F., \&  Böhme, H. J. (2020). Training a broad-coverage German sentiment classification model for dialog systems. In {\em Proceedings of The 12th Language Resources and Evaluation Conference,} 1627-1632, Marseille: France.

    \bibitem[Jalonen \& Takala(2018)]{jalonentakala18}
    Jalonen, J., \& Takala, T. (2018). Debtors' Ethical Perceptions of the Debt Collection Process. {\em Electronic Journal of Business Ethics and Organization Studies, 23}(1), 14-19.

    \bibitem[Klarna(2021)]{klarna21}
    Klarna (2021, May 3). {\em Weltweiter Vergleich zeigt: Deutschland hinkt trotz Corona bei digitalen Bezahlmethoden hinterher.} Klarna Bank AB. \url{https://www.klarna.com/international/press/weltweiter-vergleich-zeigt-deutschland-hinkt-trotz-corona-bei-digitalen-bezahlmethoden-hinterher/} 

    \bibitem[Kolnhofer-Derecskei et al.(2017)]{kolnhoferetal17}
    Kolnhofer-Derecskei, A., Reicher, R. Z., \& Szeghegyi, A. (2017). The X and Y generations’ characteristics comparison. {\em Acta Polytechnica Hungarica, 14}(8), 107-125.

    \bibitem[Kusa et al.(2020)]{kusaetal20}
    Kusa, A., Zauskova, A., \& Cabyova, L. (2020). Effect of marketing communication on consumer preferences and purchasing decisions. {\em Ad Alta: Journal of Interdisciplinary Research, 10}(1), 150-155.
        
    \bibitem[Lipiäinen(2014)]{lipiäinen14}
    Lipiäinen, H. (2014). {\em Digitization of the Communication and its Implications for Marketing.} (Doctoral Dissertation). University of Jyväskylä: Jyväskylä.

    \bibitem[Lorenz-Spreen et al.(2019)]{lorenzetal19}
    Lorenz-Spreen, P., Mønsted, B. M., Hövel, P., \& Lehmann, S. (2019). Accelerating dynamics of collective attention. {\em Nature Communications, 10}(1), 1-9. \url{https://doi.org/10.1038/s41467-019-09311-w}
    
    \bibitem[Madell \& Muncer(2007)]{madellmuncer07}
    Madell, D. E., \& Muncer, S. J. (2007). Control over social interactions: an important reason for young people's use of the Internet and mobile phones for communication?. {\em Cyberpsychology \& behavior, 10}(1), 137-140. \url{https://doi.org/10.1089/cpb.2006.9980}

    \bibitem[McCrae et al.(1999)]{mccraeetal99}
    McCrae, R. R., Costa, P. T., de Lima, M. P., Simões, A., Ostendorf, F., Angleitner, A., ... \& Piedmont, R. L. (1999). Age differences in personality across the adult life span: parallels in five cultures. {\em Developmental psychology, 35}(2), 466-477. \url{https://doi.org/10.1037/0012-1649.35.2.466}

   \bibitem[Meyer(2022)]{meyer22}
    Meyer, B. (2022, November 16). {\em The best time to send emails (2022 research).} Omnised. \url{https://www.omnisend.com/blog/best-time-to-send-email/}

    \bibitem[Michie et al.(2011)]{michieetal11}
    Michie, S., Van Stralen, M. M., \& West, R. (2011). The behaviour change wheel: A new method for characterising and designing behaviour change interventions. {\em Implementation Science, 6}(1), 1-11.

   \bibitem[Microsoft(2015)]{microsoft15}
    Microsoft (2015). {\em Attention spans.} Consumer Insights, Microsoft Canada. \url{https://dl.motamem.org/microsoft-attention-spans-research-report.pdf}
    
    \bibitem[Mihart(2012)]{mihart12}
    Mihart, C. (2012). Impact of integrated marketing communication on consumer behaviour: Effects on consumer decision-making process. {\em International Journal of Marketing Studies, 4}(2), 121-129.

    \bibitem[Mondres(2019)]{mondres19}
    Mondres, T. (2019). How Generation Z is changing financial services. {\em American Bankers Association. ABA Banking Journal, 111}(1), 24-28. 

    \bibitem[Muench \& Baumel(2017)]{muenchbaumel17}
    Muench, F., \& Baumel, A. (2017). More than a text message: dismantling digital triggers to curate behavior change in patient-centered health interventions. {\em Journal of medical Internet research, 19}(5), 1-14.

    \bibitem[Pierce(2009)]{pierce09}
    Pierce, T. (2009). Social anxiety and technology: Face-to-face communication versus technological communication among teens. {\em Computers in Human Behavior, 25}(6), 1367-1372.

    \bibitem[Rechtsdienstleistungsgesetz(2021)]{recht21}
    Rechtsdienstleistungsgesetz, §13a Darlegungs- und Informationspflichten bei Inkassodienstleistungen gegenüber Privatpersonen (2021). \url{https://dejure.org/gesetze/RDG/13a.html}

    \bibitem[Statistisches Bundesamt(2019)]{bundesamt19}
    Statistisches Bundesamt (2019, May 28). {\em Private Überschuldung: Starke Unterschiede zwischen Jung und Alt.} \url{https://www.destatis.de/DE/Presse/Pressemitteilungen/2019/05/PD19_199_635.html}

    \bibitem[Statistisches Bundesamt(2023)]{bundesamt23}
    Statistisches Bundesamt (2023, January 12). {\em P28\% der Überschuldeten hatten 2021 Schulden bei Onlinehändler.} \url{https://www.destatis.de/DE/Presse/Pressemitteilungen/2023/01/PD23_N001_63511.html}

    \bibitem[Taneja et al.(2018)]{tanejaetal18}
    Taneja, H., Wu, A. X., \& Edgerly, S. (2018). Rethinking the generational gap in online news use: An infrastructural perspective. {\em New Media \& Society, 20}(5), 1792-1812. \url{https://doi.org/10.1177/1461444817707348}

    \bibitem[Taylor et al.(2016)]{tayloretal16}
    Taylor, M. J., Arriscado, D., Vlaev, I., Taylor, D., Gately, P., \& Darzi, A. (2016). Measuring perceived exercise capability and investigating its relationship with childhood obesity: a feasibility study. {\em International Journal of Obesity, 40}(1), 34-38.

    \bibitem[Twilio(2019)]{twilio19}
    Twilio, Inc. (2019, November 7). {\em Twilio Study Shows Consumers Aren’t Paying Attention to Brands’ Social Media or Mobile Apps — They Prefer Email or Text Instead.} Twilio, Inc. \url{https://www.twilio.com/press/releases/twilio-study-shows-consumers-arent-paying-attention-to-brands-social-media-or-mobile-apps-they-prefer-email-or-text-instead}
    
    \bibitem[Willmott et al.(2021)]{willmottetal21}
    Willmott, T. J., Pang, B., \& Rundle-Thiele, S. (2021). Capability, opportunity, and motivation: an across contexts empirical examination of the COM-B model. {\em BMC Public Health, 21}(1), 1-17.

    \bibitem[Yuan et al.(2016)]{yuanetal16}
    Yuan, S., Hussain, S. A., Hales, K. D., \& Cotten, S. R. (2016). What do they like? Communication preferences and patterns of older adults in the United States: The role of technology. {\em Educational Gerontology, 42}(3), 163-174.

    \bibitem[Zerquera et al.(2016)]{zerqueraetal16}
    Zerquera, D. D., McGowan, B. L., \& Ferguson, T. L. (2016). Yes, No, Maybe So: College Students' Attitudes Regarding Debt. {\em Journal of College Student Development, 57}(5), 609-613.

 \end{thebibliography}

\newpage

\section*{Appendix}
\setcounter{table}{0}
\renewcommand{\thetable}{A\arabic{table}}

\subsection*{Payment and Communication Preferences}
\begin{table}[ht]
	\caption{Prediction of ranked preferences for payment methods using ordered logistic regression.}
	\centering
	\begin{tabular}{lccccc}
		\toprule
		   & Manual transfer & \makecell{Manual transfer \\ via Klarna} & Credit card & Paypal & Apple Pay \\
		\midrule
        Age & 1.03*** & 0.99 & 1.01* & 0.97*** & 0.97*** \\
          & [1.02, 1.04] & [0.98, 1.01] & [1.00, 1.02] & [0.96, 0.98] & [0.96, 0.98] \\
        Gender female & 0.93 & 1.77*** & 0.61*** & 1.09 & 0.81 \\
          & [0.72, 1.20] & [1.38, 2.28] & [0.47, 0.78] & [0.84, 1.41] & [0.62, 1.06] \\
        Income & 1.00 & 1.00 & 1.00* & 1.00* & 1.00** \\
          & [1.00, 1.00] & [1.00, 1.00] & [1.00, 1.00] & [1.00, 1.00] & [1.00, 1.00] \\
        Urban residency & 1.03 & 0.89 & 1.25** & 1.00 & 1.28** \\
          & [0.87, 1.22] & [0.76, 1.05] & [1.06, 1.48] & [0.84, 1.18] & [1.07, 1.53] \\
        \bottomrule
        Observations & 872 & 872 & 872 & 872 & 872 \\
		\bottomrule
	\end{tabular}
        \small
        \begin{itemize}
        \item[] \textit{Note.} Odds ratios are reported. Values in brackets are 95\% confidence intervals, * \textit{p} < .05, ** \textit{p} < .01, *** \textit{p} < .001. \\
        \end{itemize}
	\label{tab:regression_payments}
\end{table}

\begin{table}[ht]
	\caption{Prediction of ranked preferences for communication channels using ordered logistic regression.}
	\centering
	\begin{tabular}{lccccc}
		\toprule
		   & Email & Letter & Whatsapp & SMS & Call \\
		\midrule
        Age & 0.99 & 1.04*** & 0.99 & 0.99 & 1.00 \\
          & [0.98,1.00] & [1.02,1.05] & [0.98,1.00] & [0.98,1.00] & [0.99,1.01] \\
        Gender female & 1.05 & 1.17 & 0.90 & 1.11 & 0.90 \\
          & [0.80,1.37] & [0.90,1.54] & [0.70,1.17] & [0.86,1.43] & [0.70,1.17] \\
        Income & 1.00 & 1.00 & 1.00*** & 1.00 & 1.00 \\
          & [1.00, 1.00] & [1.00, 1.00] & [1.00, 1.00] & [1.00, 1.00] & [1.00, 1.00] \\
        Urban residency & 0.98 & 1.14 & 0.98 & 1.32** & 1.01 \\
          & [0.82,1.16] & [0.95,1.37] & [0.82,1.16] & [1.11,1.57] & [0.85,1.20]] \\
        \bottomrule
        Observations & 872 & 872 & 872 & 872 & 872 \\
		\bottomrule
	\end{tabular}
        \small
        \begin{itemize}
        \item[] \textit{Note.} Odds ratios are reported. Values in brackets are 95\% confidence intervals, * \textit{p} < .05, ** \textit{p} < .01, *** \textit{p} < .001. \\
        \end{itemize}
	\label{tab:regression_channels}
\end{table}

\begin{table}[ht]
	\caption{Prediction of ranked preferences for communication channels using ordered logistic regression.}
	\centering
	\begin{tabular}{lcccc}
		\toprule
		   & Morning & Noon & Afternoon & Evening \\
		\midrule
        Age & 1.01 & 1.01 & 0.99* & 0.99 \\
          & [0.99,1.02] & [0.99,1.02] & [0.98,1.00] & [0.98,1.01] \\
        Gender female & 1.17 & 1.28 & 0.71* & 0.82 \\
          & [0.88,1.54] & [0.97,1.69] & [0.54,0.94] & [0.62,1.08] \\
        Income & 1.00** & 1.00 & 1.00 & 1.00 \\
          & [1.00, 1.00] & [1.00, 1.00] & [1.00, 1.00] & [1.00, 1.00] \\
        Urban residency & 1.09 & 1.04 & 0.93 & 0.79* \\
          & [0.90,1.31] & [0.87,1.25] & [0.77,1.12] & [0.66,0.95]\\
        \bottomrule
        Observations & 754 & 763 & 777 & 756 \\
		\bottomrule
	\end{tabular}
        \small
        \begin{itemize}
        \item[] \textit{Note.} Odds ratios are reported. Values in brackets are 95\% confidence intervals, * \textit{p} < .05, ** \textit{p} < .01, *** \textit{p} < .001. \\
        \end{itemize}
	\label{tab:regression_time}
\end{table}

\setcounter{figure}{0}
\renewcommand{\thefigure}{A\arabic{figure}}

\subsection*{Geographical Differences}

\begin{figure}[ht]
	\centering
	\caption{Geographical differences across Germany in terms of payment method preferences. Darker colors within each graph represent stronger preferences for a specific payment method.} 
	\includegraphics[width=14cm, height=16cm]{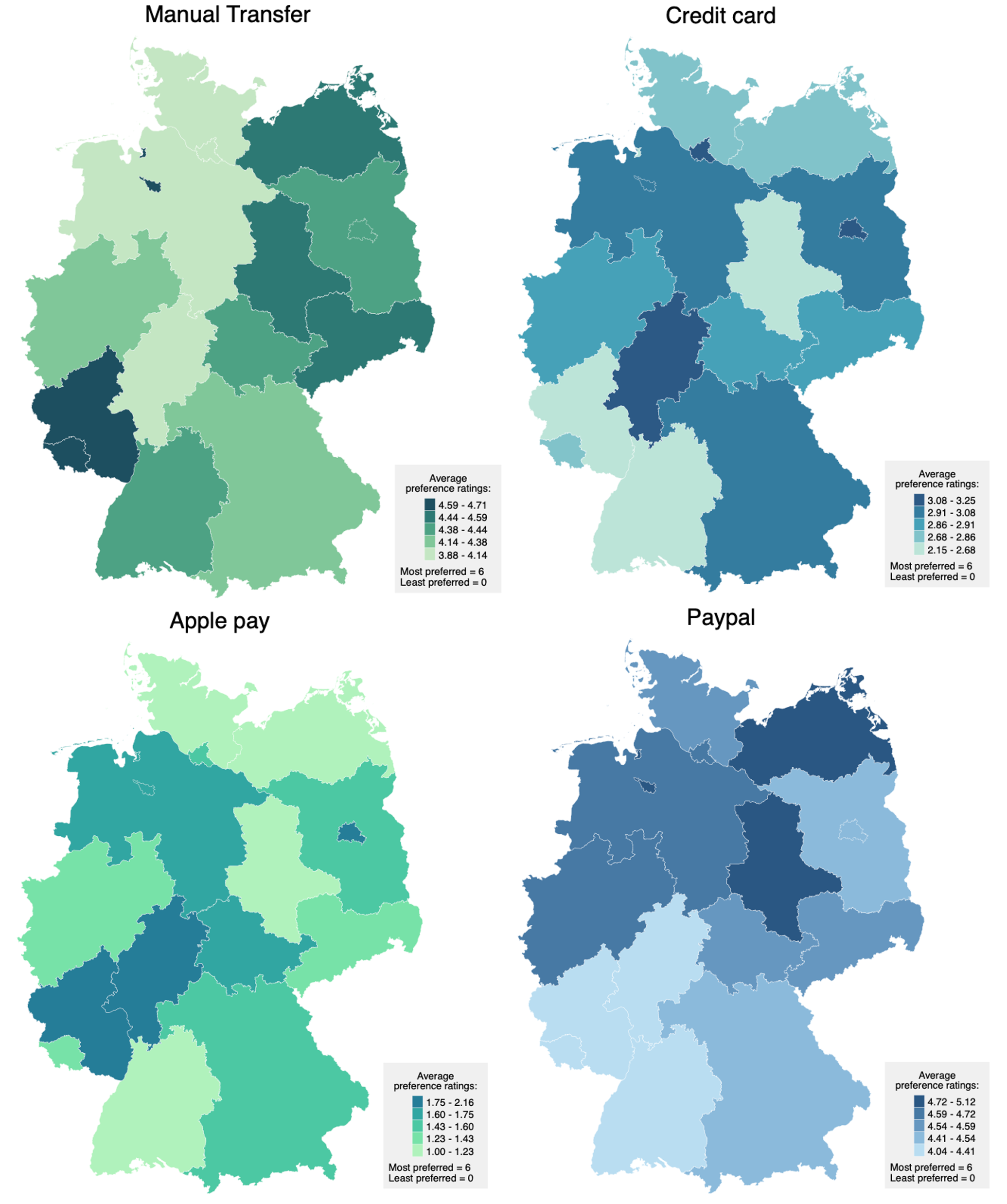}
	\label{fig:geography_payments}
\end{figure}

In an explorative analysis, we investigated geographical differences in terms of payment and communication preferences across Germany. To do this, we display average rankings of payment methods on the level of federal states (see Figure~\ref{fig:geography_payments}). This analysis enables companies to adapt their payment method options and communication strategies for each individual consumer based on their location. 

\begin{figure}[ht]
	\centering
	\caption{Geographical differences across Germany in terms of communication channel preferences. Darker colors within each graph represent stronger preferences for a specific communication channel.} 
	\includegraphics[width=13cm, height=22cm]{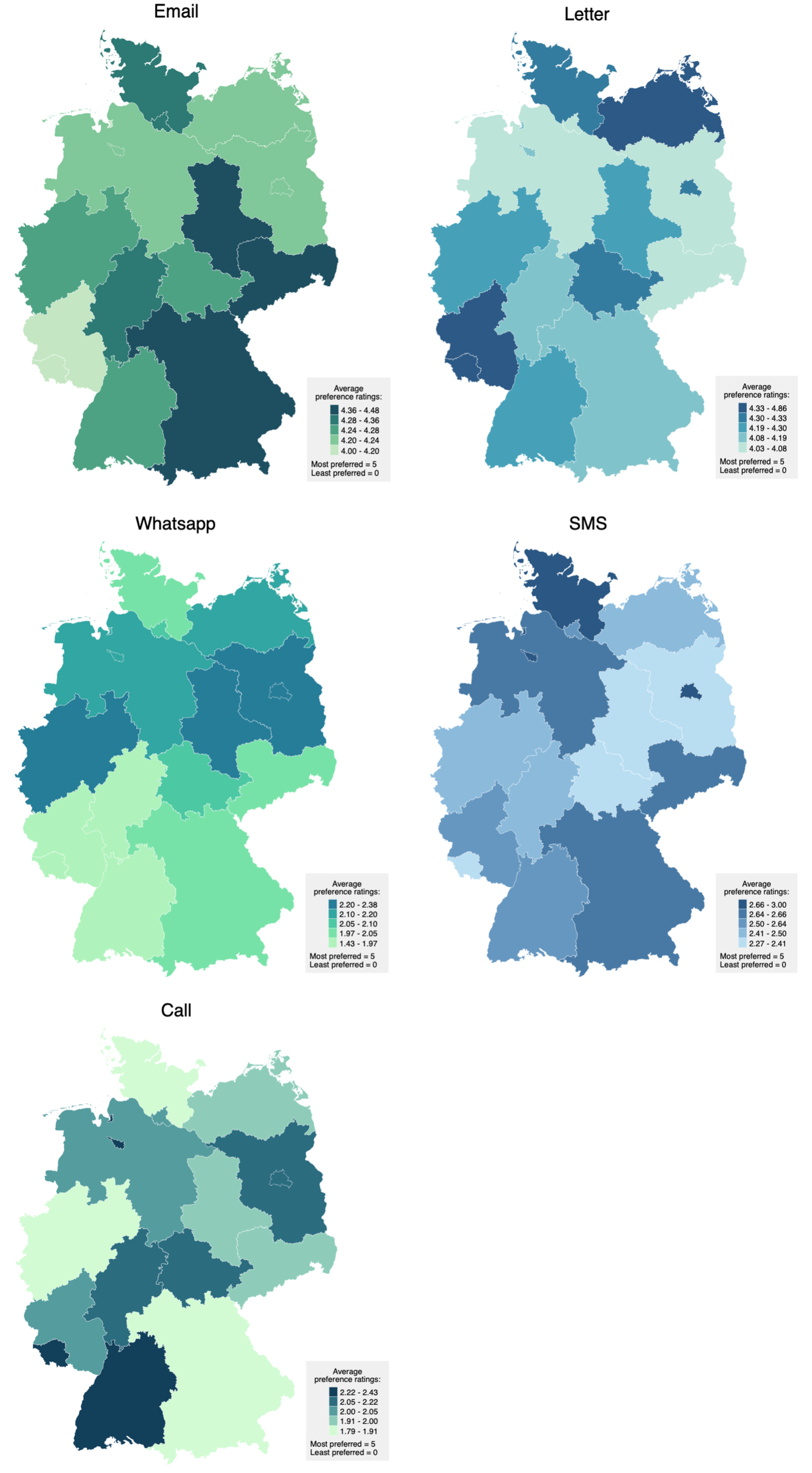}
	\label{fig:geography_channels}
\end{figure}

\begin{figure}[ht]
	\centering
	\caption{Geographical differences across Germany in terms of communication time preferences. Darker colors within each graph represent stronger preferences for a specific time of day.} 
	\includegraphics[width=14cm, height=16cm]{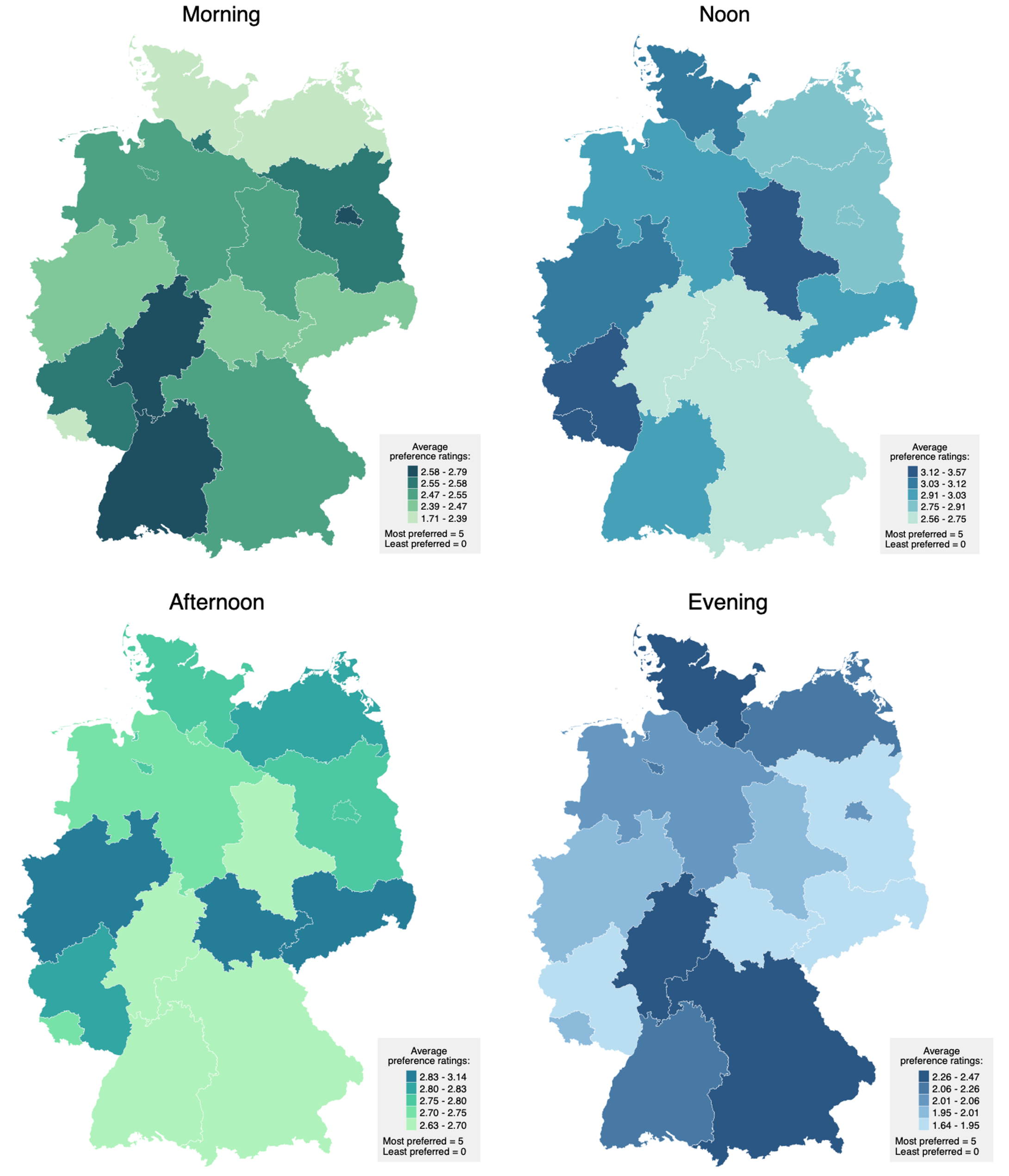}
	\label{fig:geography_timing}
\end{figure}

We conducted a similar explorative analysis for communication channels (see Figure~\ref{fig:geography_channels}) and preference for communication time (see Figure~\ref{fig:geography_timing}). 

\FloatBarrier
\newpage
\subsection*{Attitudes}

\begin{table}[ht]
	\caption{Prediction of openness to talk about debt using ordered logistic regression.}
	\centering
	\begin{tabular}{lcccc}
		\toprule
		   & Openness \\
		\midrule
        Age & 1.01* \\
          & [1.00,1.02] \\
        Gender female & 1.25 \\
          & [0.98,1.61] \\
        \bottomrule
        Observations & 888 \\
		\bottomrule
	\end{tabular}
        \small
        \begin{itemize}
        \item[]  \textit{Note.} Odds ratios are reported. Values in brackets are 95\% confidence intervals, * \textit{p} < .05, ** \textit{p} < .01, *** \textit{p} < .001. \\
        \end{itemize}
	\label{tab:regression_openness}
\end{table}

\begin{table}[ht]
	\caption{Attitudes: Prediction of adjectives' sentiments using ordered logistic regression.}
	\centering
	\begin{tabular}{lcccc}
		\toprule
		   & Towards DCAs & \makecell{Towards individuals \\ contacted by DCAs} & \makecell{Towards individuals \\ contacted the most by DCAs} \\
		\midrule
        Age & 0.99 & 0.98 & 1.00 \\
          & [0.98,1.00] & [0.97,1.00] & [0.99,1.01] \\
        Gender female & 1.45*** & 0.87 & 1.03 \\
          & [1.23,1.73] & [0.65,1.15] & [0.81,1.30] \\
        Experience with DCA & 0.42*** & 1.66*** & 1.65***  \\
          & [0.34, 0.51] & [1.26, 2.21] & [1.30, 2.10] \\
        \bottomrule
        Observations & 2569 & 2361 & 2258 \\
		\bottomrule
	\end{tabular}
        \small
        \begin{itemize}
        \item[]  \textit{Note.} Odds ratios are reported. Values in brackets are 95\% confidence intervals, * \textit{p} < .05, ** \textit{p} < .01, *** \textit{p} < .001. \\
        \end{itemize}
	\label{tab:regression_attitudes}
\end{table}

\begin{table}[ht]
	\caption{Adjectives associated with Debt Collection Agencies.}
	\centering
	\begin{tabular}{llll}
		\toprule
		Original response & Translation & Sentiment & n \\
		\midrule
		unseriös & illegitimate & negative & 107 \\
        notwendig & necessary & positive & 106 \\
        unangenehm & unpleasant & negative & 86 \\
        teuer & expensive & negative & 58 \\
        fordernd & demanding & negative & 55 \\
        hilfreich & helpful & positive & 47 \\
        nützlich & useful & positive & 46 \\
        streng & strict & negative & 46 \\
        wichtig & important & positive & 45 \\
        nervig & annoying & negative & 43 \\
        aufdringlich & intrusive & negative & 41 \\
        hart & hard & negative & 41 \\
        unfreundlich & unfriendly & negative & 38 \\
        einschüchternd & intimidating & negative & 37 \\
        gut & good & positive & 37 \\
        bedrohlich & threatening & negative & 36 \\
        schlecht & bad & negative & 34 \\
        negative & negative & negative & 32 \\
        aggressiv & aggressive & negative & 31 \\
        gierig & greedy & negative & 31 \\
        sinnvoll & sensible & positive & 28 \\
        nötig & needed & positive & 26 \\
        seriös & legitimate & positive & 25 \\
        konsequent & resolute & neutral & 24 \\
        abzockend & rip-off & negative & 23 \\
        beängstigend & frightening & negative & 23 \\
        gerecht & just & positive & 22 \\
        lästig & annoying & negative & 21 \\
        zwielichtig & shady & negative & 21 \\
        gefährlich & dangerous & negative & 20 \\
        schnell & fast & neutral & 20 \\
        gemein & mean & negative & 19 \\
        brutal & brutal & negative & 18 \\
        dubious & dubious & negative & 18 \\
        böse & evil & negative & 17 \\
        hartnäckig & stubborn & negative & 17 \\
        skrupellos & ruthless & negative & 17 \\
        unfair & unfair & negative & 17 \\
        ängstlich & anxious & negative & 16 \\
        durchsetzungsfähig & assertive & neutral & 16 \\
        unbeliebt & unpopular & negative & 16 \\
        kriminell & criminal & negative & 15 \\
		\bottomrule
	\end{tabular}
          \small
          \begin{itemize}
          \item[] \hspace*{25mm} \textit{Note.} Total responses \textit{n} = 2600, most common responses \textit{n} = 1436.
          \end{itemize}
	\label{tab:adjectives_1}
\end{table}

\begin{table}[ht]
	\caption{Adjectives associated with individuals who get contacted by Debt Collection Agencies.}
	\centering
	\begin{tabular}{llll}
		\toprule
		Original response & Translation & Sentiment & n \\
		\midrule
        arm & poor & negative & 252 \\
        unzuverlässig & unreliable & negative & 96 \\
        unorganisiert & unorganized & negative & 65 \\
        verschuldet & indebted & negative & 61 \\
        traurig & sad & negative & 58 \\
        vergesslich & forgetful & negative & 57 \\
        zahlungsunfähig & illiquid & negative & 48 \\
        bemitleidenswert & pitiful & negative & 47 \\
        hilflos & helpless & negative & 42 \\
        unglücklich & unhappy & negative & 40 \\
        pleite & broke & negative & 36 \\
        schuldig & guilty & negative & 34 \\
        verzweifelt & desperate & negative & 32 \\
        überfordert & overwhelmed & negative & 31 \\
        dumm & stupid & negative & 26 \\
        faul & lazy & negative & 26 \\
        hilfsbedürftig & needy & negative & 25 \\
        nachlässig & negligent & negative & 24 \\
        problematisch & problematic & negative & 24 \\
        verschwenderisch & wasteful & negative & 23 \\
        chaotisch & chaotic & negative & 22 \\
        schlecht & bad & negative & 22 \\
        ängstlich & anxious & negative & 21 \\
        naiv & naive & negative & 21 \\
        verantwortungslos & irresponsible & negative & 20 \\
        normal & normal & neutral & 19 \\
        bedauernswert & pitiable & negative & 18 \\
        leichtsinnig & reckless & negative & 18 \\
        neutral	& neutral & neutral & 18 \\
        unsicher & insecure & negative & 17 \\
        unverschuldet & innocent & positive	& 16 \\
        säumig & defaulting & negative & 15 \\
        verplant & muddled & negative & 15 \\
		\bottomrule
	\end{tabular}
          \small
          \begin{itemize}
          \item[] \hspace*{25mm} \textit{Note.} Total responses \textit{n} = 2393, most common responses \textit{n} = 1289.
          \end{itemize}
	\label{tab:adjectives_2}
\end{table}

\begin{figure}[ht]
	\centering
	\caption{Attitudes towards people contacted most by Debt Collection Agencies expressed via adjectives. Larger font sizes indicate more frequent occurrences (minimum frequency displayed is 15). Adjectives with a positive sentiment are displayed in orange, negative sentiments are displayed in blue and neutral ones are displayed in pink. The figure includes 61.6\% of the responses rated for sentiment.} 
	\includegraphics[width=10cm, height=8cm]{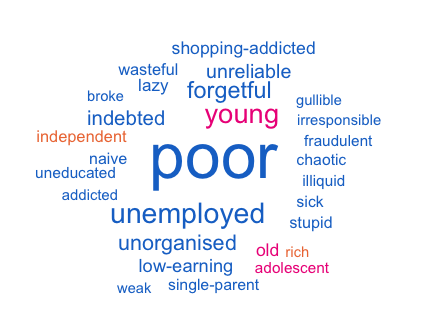}
	\label{fig:attitudes_3}
\end{figure}

When asked about characteristics of people who get contacted by DCAs the most, participants predominantly provided adjectives of negative sentiment (84.8\%). However, an increase in neutral responses was observed (10.3\%) while the number of positive sentiments decreased even further (4.9\%) in comparison to the responses for individuals who get contacted by DCAs in general. For a more detailed overview of adjectives associated with people who get contacted by DCAs frequently, see Figure~\ref{fig:attitudes_3}. Again, no significant effect of age $(OR = 1.0, z = 0.34, p = .73)$ or gender $(OR = 1.04, z = 0.37, p = .70)$ could be observed. Prior experience with DCAs was associated with significantly more positive attitudes $(OR = 1.63, z = 3.99, p < .001)$

\begin{table}[ht]
	\caption{Adjectives associated with individuals who get contacted the most by Debt Collection Agencies.}
	\centering
	\begin{tabular}{llll}
		\toprule
		Original response & Translation & Sentiment & n \\
		\midrule
        arm & poor & negative & 401 \\
        arbeitslos & unemployed & negative & 123 \\
        jung & young & neutral & 114 \\
        vergesslich & forgetful & negative & 83 \\
        unorganisiert & unorganized & negative & 69 \\
        verschuldet & indebted & negative & 60 \\
        unzuverlässig & unreliable & negative & 54 \\
        geringverdienend & low-earning & negative & 43 \\
        alt & old & neutral & 38 \\
        kaufsüchtig & shopping-addicted & negative & 38 \\
        faul & lazy & negative & 33 \\
        verschwenderisch & wasteful & negative & 28 \\
        selbstständig & independent & positive & 28 \\
        dumm & stupid & negative & 27 \\
        naiv & naive & negative & 25 \\
        krank & sick & negative & 24 \\
        alleinerziehend & single-parent & negative & 23 \\
        chaotisch & chaotic & negative & 23 \\
        ungebildet & uneducated & negative & 21 \\
        zahlungsunfähig & illiquid & negative & 21 \\
        betrügerisch & fraudulent & negative & 20 \\
        jugendlich & adolescent & neutral & 20 \\
        süchtig & addicted & negative & 17 \\
        reich & rich & positive & 16 \\
        verantwortungslos & irresponsible & negative & 16 \\
        leichtgläubig & gullible & negative & 15 \\
        pleite & broke & negative & 15 \\
        schwach & weak & negative & 15 \\
		\bottomrule
	\end{tabular}
    \small
    \begin{itemize}
    \item[] \hspace*{25mm} \textit{Note.} Total responses \textit{n} = 2287, most common responses \textit{n} = 1410.
    \end{itemize}
	\label{tab:adjectives_3}
\end{table}

\FloatBarrier
\subsection*{Questionnaire (Translated version)}

\textbf{Purpose of the study} \\
The purpose of this survey is to generate a deeper understanding of young consumers' payment behavior. In particular, it focuses on the use of digital options. Furthermore, the attitude towards debt collection agencies and debtors is investigated. Therefore, please read carefully the description of the activities of a debt collection agency in the following section.

\textbf{What does a debt collection agency do? } \\
All companies, from small tradesmen to large corporations, have the problem that some of their invoices are not paid within the agreed time frame. If a company is also unable to achieve payment within the framework of its own dunning procedure, because they cannot contact the defaulting consumer or the payment remains outstanding, they engage specialized service providers, the debt collection agency. A debt collection agency now tries to contact the customer and find an amicable solution to the outstanding debt. If no payment agreement is reached in the process, collection agencies are also allowed to also initiate legal steps, such as the judicial dunning process and enforcement. 

\textbf{What is your attitude towards debt collection agencies/ What do you think about debt collection agencies in general?} \\
Please provide three adjectives. 

\textbf{Would you tell a friend if you were contacted by a collection agency?} \\	
Please select one answer: \\
Yes, I would talk to my friends about it \\
Yes, but only to my closest friend(s) \\
No, I would rather keep it to myself \\
No, I don't think the topic is important among friends 

\textbf{What do you think about people who are/were in a debt collection process?} \\
Please provide three adjectives. 
						
\textbf{Which people do you think are most often contacted by debt collection agencies?} \\	
Please provide three adjectives. 
						
\textbf{Have you ever received a dunning letter?} \\
Yes, once \\
Yes, several times \\
No \\
I do not know 
				
\textbf{Have you ever been contacted by a debt collection agency?}  \\
Yes, once  \\
Yes, several times  \\
No  \\
I do not know  

If "Yes" was previously selected:				
\textbf{How would you describe your experience with the collection agency?}  \\
Positive \\
Negative \\
Neutral \\
Other experience:

\textbf{Why did you not pay the invoice?}  \\
I had forgotten \\
I was not at home for a long time and therefore did not see the invoice \\
I could not remember what I ordered \\
I did not have enough money \\
I was dissatisfied with the service or the product \\
I returned the product too late \\
Other reason: 

\textbf{How did you make your payment?}  \\
Bank transfer (manually) \\
Bank transfer (via Klarna) \\
Credit card \\
Paypal \\
Apple Pay \\
Cash payment (payment of the invoice in cash e.g. in the supermarket) \\
Other payment method: 
				
\textbf{Do you open every email in your inbox? If not, which ones do you not open and why?
Several answers can be selected:}  \\
Yes, I open every email \\
If I already get enough information in the subject line, I don't open the email \\
If I do not know the sender, I do not open the email \\
If there are grammatical errors in the subject line, I do not open the email \\
If the subject line is written in capital letters, I do not open the email \\
If the email has emojis in the subject line, I don't open the email \\
Other reason why I don't open an email:

\textbf{Which payment method do you prefer to use to pay your invoice? }  \\
Please rank the payment methods (1 = I prefer/frequently use it to pay, 6 = I dislike/rarely use it to pay): \\
Bank transfer (manual) \\
Bank transfer (via Klarna) \\
Credit card \\
Paypal \\
Apple Pay \\
“Barzahlen” (payment of the invoice in cash, e.g. in the supermarket). 
		 	 	 		
\textbf{Example scenario:}  \\
Imagine the following scenario: You bought a sweater through an online platform and did not pay the invoice. You have not responded to the payment reminders and now your case is being handled by a collection agency. Due to the delay in payment, further costs have now been incurred. The collection agency is now responsible for informing you that the invoice is still open and requesting you to make a payment. 

\textbf{Via which channels would you prefer to receive the payment reminder?}  \\
Please rank the channels (1 = I would prefer to receive the reminder through this channel, 5 = I would be very reluctant to be contacted through this channel): \\
Email \\
Letter \\
Whatsapp \\
SMS \\
Call \\
Other channel: 

\textbf{At what time of day would you prefer to receive a payment reminder via email?}  \\
Please rank the times (1 = I would most like to receive the reminder at this time, 4 = I would be very reluctant to receive the reminder at this time): \\
In the morning between 6:00 and 10:00 a.m. \\
At noon between 10:00 and 14:00 \\
In the afternoon between 14:00 and 18:00 \\
In the evening between 18:00 and 22:00 \\
Other time: 

\textbf{If it would be difficult for you to pay the whole amount at the same time, what type of solutions would be interesting to you?}  \\
You can choose several answers: \\
Payment pause (postponing the payment for 4 weeks) \\
Installment payment (payment of the total amount in smaller amounts over a longer period of time) \\
Other solution offer: 
		
\textbf{Please indicate your age:}

\textbf{Please indicate your gender:} \\
Female \\
Male \\
Divers

\textbf{Please enter the zip code of your place of residence:}

\textbf{Please indicate your monthly disposable income (net, in €):}

\textbf{Please indicate your rent per square meter (in €):} \\
Rent per square meter in € \\
I do not live for rent
				
The survey is now finished.\footnote{Participants were additionally asked to rate the content of different payment reminders. Due to confidentiality reasons, the messages are not included here.}
						
Thank you for your participation!

\subsection*{Questionnaire (original language: German)}

\textbf{Ziel der Studie} \\
In der vorliegenden Befragung geht es darum, ein tieferes Verstaendnis ueber das Zahlungsverhalten von jungen Menschen zu generieren. Es geht dabei insbesondere um die Nutzung digitaler Angebote. Weiterhin wird die Einstellung gegenueber Inkasso-Unternehmen und Schuldnern untersucht. Bitte lesen Sie sich daher die Beschreibung der Taetigkeit eines Inkasso-Unternehmens im folgenden Abschnitt aufmerksam durch.
						
\textbf{Was macht ein Inkassounternehmen?}	\\
Alle Unternehmen, vom kleinen Handwerker bis zum Grossunternehmen, haben das Problem, dass ein Teil ihrer Rechnungen nicht innerhalb der vereinbarten Frist bezahlt wird. Wenn ein Unternehmen im Rahmen seines eigenen Mahnverfahrens auch keine Zahlung erreichen kann, weil sie den saeumigen Konsumenten nicht kontaktieren koennen oder die Zahlung weiterhin offen ist, beauftragen sie spezialisierte Dienstleister, die Inkassounternehmen. Ein Inkassounternehmen versucht nun mit dem Kunden in Kontakt zu treten und eine einvernehmliche Loesung fuer die ausstehende Forderung zu finden. Sollte dabei keine Zahlungsvereinbarung erreicht werden, duerfen Inkassounternehmen auch gerichtliche Schritte, wie z.B. das gerichtliche Mahnverfahren und die Zwangsvollstreckung, einleiten. 
	
\textbf{Was ist Ihre Einstellung zu Inkasso-Unternehmen / Was denken Sie allgemein ueber Inkasso-Unternehmen?} \\
Bitte geben Sie drei Adjektive an. 

\textbf{Wuerden Sie einem Freund davon erzaehlen, wenn Sie von einem Inkasso-Unternehmen kontaktiert werden?} \\
Bitte waehlen Sie eine Antwort aus: \\
Ja, ich wuerde mich mit meinen Freunden darueber austauschen \\
Ja, aber nur mit meinem engsten Freund / meiner engsten Freundin \\
Nein, ich wuerde es eher für mich behalten \\
Nein, ich halte das Thema unter Freunden für unwichtig 
						
\textbf{Was denken Sie ueber Menschen, die sich in einem Inkasso-Verfahren befinden/befanden?} \\
Bitte geben Sie drei Adjektive an. 	
						
\textbf{Welche Menschen glauben Sie werden am haeufigsten von Inkasso-Unternehmen kontaktiert?}	\\
Bitte geben Sie drei Adjektive an. 
						
\textbf{Haben Sie schon einmal eine Mahnung erhalten?} \\
Ja, einmal \\
Ja, mehrmals \\
Nein \\
Weiß ich nicht 
				
\textbf{Wurden Sie schon einmal von einem Inkasso-Unternehmen kontaktiert?} \\
Ja, einmal \\
Ja, mehrmals \\
Nein \\
Weiß ich nicht 

\textbf{Wenn zuvor “Ja” ausgewählt wurde:}	\\
Wie wuerden Sie Ihre Erfahrung mit dem Inkasso-Unternehmen beschreiben? \\
Positiv \\
Negativ \\ 
Neutral \\
Andere Erfahrung: 
				
\textbf{Warum haben Sie die Rechnung nicht gezahlt?} \\
Ich hatte es vergessen \\
Ich war laenger nicht zu Hause und habe deshalb die Rechnung nicht gesehen \\
Ich konnte mich nicht mehr erinnern, was ich bestellt habe \\
Ich hatte nicht mehr genug Geld \\
Ich war unzufrieden mit der Leistung bzw. dem Produkt \\
Ich habe das Produkt zu spaet retourniert \\
Anderer Grund:

\textbf{Wie haben Sie Ihre Zahlung geleistet?} \\
Ueberweisung (manuell) \\
Ueberweisung (via Klarna) \\
Kreditkarte \\
Paypal \\
Apple Pay \\
Barzahlen (Zahlung der Rechnung in Bargeld z.B. im Supermarkt) \\
Weitere Zahlungsmethode:
				
\textbf{Oeffnen Sie jede Email in Ihrem Posteingang? Wenn nicht, welche oeffnen Sie nicht und weshalb?} \\
Es koennen mehrere Antworten ausgewaehlt werden: \\
Ja, ich oeffne jede Email \\
Wenn ich bereits genug Informationen in der Betreffzeile erhalte, öffne ich die Email nicht \\
Wenn ich den Absender nicht kenne, öffne ich die Email nicht \\
Wenn es grammatikalische Fehler in der Betreffzeile gibt, öffne ich die Email nicht \\
Wenn die Betreffzeile in Großbuchstaben geschrieben ist, öffne ich die Email nicht \\
Wenn die Email Emojis in der Betreffzeile hat, öffne ich die Email nicht \\
Anderer Grund warum ich eine Email nicht öffne:
	
\textbf{Welche Zahlungsmethode nutzen Sie am liebsten, um Ihre Rechnungen zu bezahlen?} \\
Bitte bringen Sie die Zahlungsmethoden in eine Rangfolge (1 = damit zahle ich am liebsten/haeufigsten, 6 = damit zahle ich ungerne/selten): \\
Ueberweisung (manuell) \\
Ueberweisung /via Klarna) \\
Kreditkarte \\
Paypal \\
Apple Pay \\
Barzahlen (Zahlung der Rechnung in Bargeld, z.B. im Supermarkt) 
 	 	 		
\textbf{Beispiel:} \\
Stellen Sie sich folgendes Szenario vor: Sie haben ueber eine Online-Plattform einen Pullover gekauft und die Rechnung nicht bezahlt. Auf die Zahlungserinnerungen haben Sie nicht reagiert und nun wird Ihr Fall von einem Inkasso-Unternehmen bearbeitet. Aufgrund der Zahlungsverzoegerung sind nun weitere Kosten entstanden. Die Aufgabe des Inkasso-Unternehmens ist es, Sie darauf hinzuweisen, dass die Rechnung weiterhin offen ist und Sie zu einer Zahlung aufzufordern.
						
\textbf{Ueber welche Kanaele wuerden Sie die Zahlungserinnerung am liebsten erhalten?} \\
Bitte bringen Sie die Kanaele in eine Rangfolge (1 = am liebsten erhalte ich die Erinnerung ueber diesen Kanal, 5 = ich wuerde nur sehr ungerne ueber diesen Kanal kontaktiert werden): \\
Email \\
Brief \\
Whatsapp \\
SMS \\
Anruf \\
Weiterer Kanal:  
 	 	 		
\textbf{Zu welcher Uhrzeit sind Sie fuer eine Zahlungserinnerung per E-Mail am aufnahmefaehigsten?} \\	
Bitte bringen Sie die Uhrzeiten in eine Rangfolge (1 = zu dieser Uhrzeit wuerde ich die Erinnerung am liebsten erhalten, 4 = ich wuerde die Erinnerung nur sehr ungerne zu dieser Uhrzeit erhalten) \\
Morgens zwischen 6:00 und 10:00 Uhr \\
Mittags zwischen 10:00 und 14:00 Uhr \\
Nachmittags zwischen 14:00 und 18:00 Uhr \\
Abends zwischen 18:00 und 22:00 Uhr \\
Weitere Uhrzeit: 
						
\textbf{Wenn es schwierig fuer Sie waere den gesamten Betrag gleichzeitig zu bezahlen, welche Loesungsangebote waeren fuer Sie interessant?} \\
Sie koennen mehrere Antworten auswaehlen. \\
Zahlpause (Aufschub der Zahlung für 4 Wochen) \\
Ratenzahlung (Zahlung des Gesamtbetrages in kleineren Betraegen ueber eine laungere Zeit hinweg) \\
Weiteres Loesungsangebot: 
				
\textbf{Bitte geben Sie Ihr Alter an:}

\textbf{Bitte geben Sie Ihr Geschlecht an:} \\
Weiblich \\
Männlich \\
Divers

\textbf{Bitte geben Sie die Postleitzahl Ihres Wohnortes an:}

\textbf{Bitte geben Sie Ihr monatliches verfügbares Einkommen an (netto, in €):}

\textbf{Bitte geben Sie an, wie hoch Ihre Miete pro Quadratmeter ist (in €):} \\
Miete pro Quadratmeter in € \\
ich wohne nicht zur Miete
		
Die Studie ist nun beendet.
						
Vielen Dank fuer Ihre Teilnahme! 

\end{document}